# Atomized spraying of liquid metal droplets on desired substrate surfaces as a generalized way for ubiquitous printed electronics

Qin Zhang [1], Yunxia Gao [1] and Jing Liu [1,2]*

1. Beijing Key Lab of CryoBiomedical Engineering and Key Lab of Cryogenics,
Technical Institute of Physics and Chemistry,
Chinese Academy of Sciences, Beijing 100190, P. R. China

2. Department of Biomedical Engineering, School of Medicine, Tsinghua University,
Beijing 100084, P. R. China

**\*Address for correspondence:**
Dr. Jing Liu
Beijing Key Lab of CryoBiomedical Engineering
  and Key Lab of Cryogenics,
Technical Institute of Physics and Chemistry,
Chinese Academy of Sciences,
Beijing 100190, P. R. China
E-mail address: jliu@mail.ipc.ac.cn
Tel. +86-10-82543765
Fax: +86-10-82543767

**Abstract:** A direct electronics printing technique through atomized spraying for patterning room temperature liquid metal droplets on desired substrate surfaces is proposed and experimentally demonstrated for the first time. This method has generalized purpose and is highly flexible and capable of fabricating electronic components on any desired target objects, with either flat or rough surfaces, made of different materials, or different orientations from 1-D to 3-D geometrical configurations. With a pre-designed mask, the liquid metal ink can be directly deposited on the substrate to form various specific patterns which lead to the rapid prototyping of electronic devices. Further, extended printing strategies were also suggested to illustrate the adaptability of the method such that the natural porous structure can be adopted to offer an alternative way of making transparent conductive film with an optical transmittance of 47% and a sheet resistance of 5.167Ω/□. Different from the former direct writing technology where large surface tension and poor adhesion between the liquid metal and the substrate often impede the flexible printing process, the liquid metal here no longer needs to be pre-oxidized to guarantee its applicability on target substrates. One critical mechanism was found as that the atomized liquid metal microdroplets can be quickly oxidized in the air due to its large specific surface area, resulting in a significant increase of the adhesive capacity and thus firm deposition of the ink to the substrate. This study established a generalized way for pervasively and directly printing electronics on various substrates which are expected to be significant in a wide spectrum of electrical engineering areas.

## 1 Introduction

In recent years, many investigators have been attracted and tremendous efforts were made to realize flexible electronics. Some typical applications within this area generally include: flexible displays [1], conformal antenna arrays, photovoltaics arrays [2], thin film transistors [3], radio-frequency identification (RFID) tags [4], flexible batteries [5,6], and electronic circuits fabricated in clothing or biomedical devices etc. [7]. The advent of printed electronics intensified the urgent need of finding innovative methods either in material part or printing technology. In this paper, a pervasive electronics fabrication technique through atomized spraying for patterning liquid metal droplets around room temperature is proposed and demonstrated for the first time. The method is extremely time-saving and needs no vacuum processes anymore as compared with conventional fabrication approaches of printed electronics such as evaporation and sputtering. Further, it can be scaled to large area flexible substrates without specialized equipment.

The method of spraying has been proven to be feasible in fabrication of some electronics, such as printed circuit board [7], light-absorbing layer on photovoltaic devices [8], flexible sensors [9], flexible display [10] and so on. However, so far almost all of the existing trials were involved in the aqueous solutions or organic base with dispersions of nano structures or metal flakes which may cause problems such as complex preparation of spraying material or lower conductivity of the ink. Siegel et al [7] applied metal flakes (commonly Ni or Ag) suspended in acrylic base and sprayed it to a substrate, producing an electrically conductive surface but with disadvantages of brittleness when bent and ferromagnetism (using Ni) that may not be desirable in a specific application. Orme et al. [11] proposed charged molten metal (63% Sn, 37% Pb) droplet deposition as a direct write technology. However, its application is limited by the requirement of heating module, toxicity of the materials and brittleness after solidification. Clearly, in order to make the spraying more generalized, a highly suitable electronic ink material is needed. Among the many promising inks, liquid metal is an interesting one which appears as excellent conductive liquid at room temperature. It has recently attracted many researchers' attention for its low melting point which allows fabrication on flexible substrates around room temperature [12-17] and the electronics making from common scale to micro size such as direct writing by brush [18], printing [12], injection [19] and stamp lithography. In this paper, we combined the atomized spraying technology with the room temperature liquid metal as printed electronics for the first time. Without dispersion of conducting materials, liquid metal can be patterned directly on any desired surfaces or substrates as flexible electronics with relatively high conductivity and no brittleness. It is expected that such ubiquitous printing technology will stimulate a series of unconventional applications in future electronic engineering area.

## 2  Principle and method

The present method of atomized spraying based liquid metal printed electronics can be divided into two basic procedures. One is that liquid metal is subject to atomized spraying and becomes micro-droplets with a thin oxide skin quickly formed on the surface. Such oxidized liquid metal film contributes significantly to adhere to the substrate surface. Another step is then the controlled deposition and adhesion of the liquid metal microdroplets on the target substrates which is presented in Fig. 1A. Since the surface tension (>400mN/m) of the liquid metal is much larger than that of ordinary liquids, it is harder for researchers to observe any splashing of liquid metal [20] which will degrade the quality of the deposition.

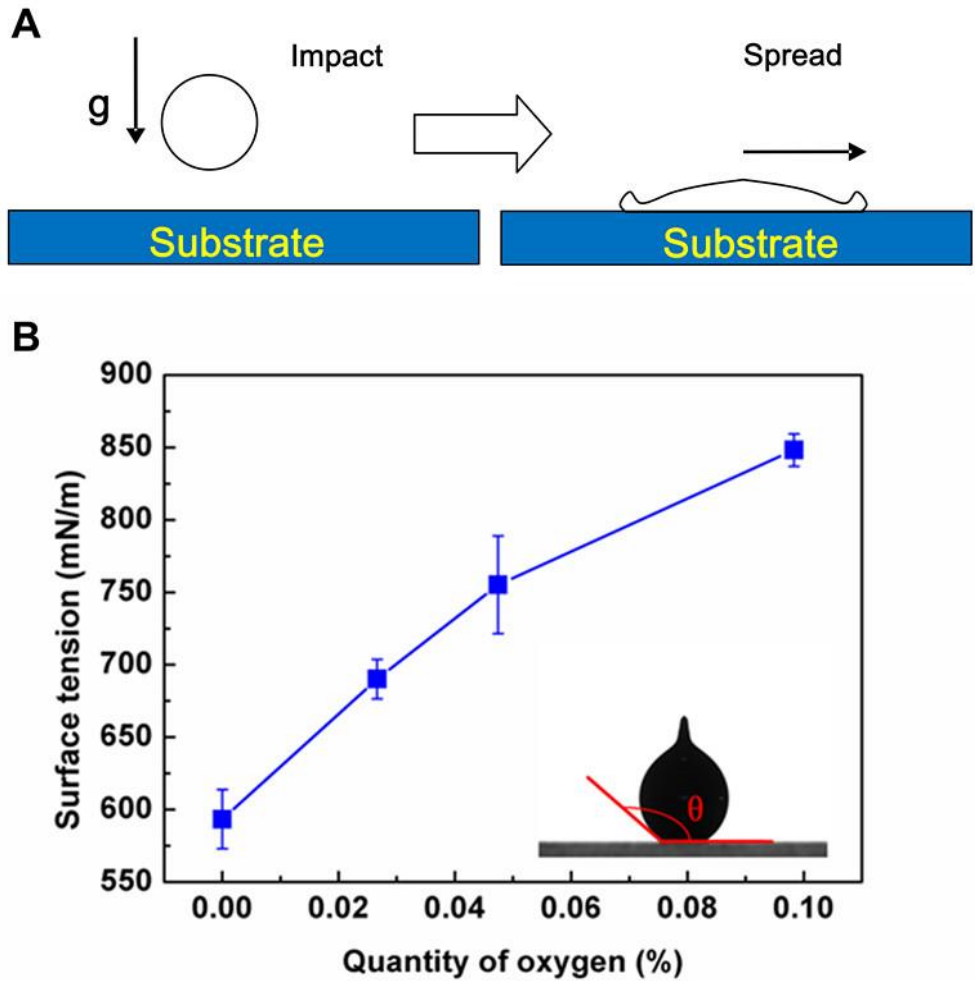

**Fig. 1** (A) Printing process and spreading impact of single droplet on a flat surface. (B) Surface tension varies with the quantity of oxygen. The insert depicts the contact angle of a liquid metal droplet.

Different from the former direct writing method [18], in the present printing mechanism, liquid metal does not need to be pre-oxidized to realize general applicability on the hard to wet substrates. As previously reported, any exposed region of the galinstan alloy to air or even as little as 0.2% volume of oxygen will cause surface oxidation of gallium and formation of a thin solid film, preventing further oxidation [21]. Compared with gallium, the oxidation of indium is very slow when exposed to oxygen [22]. Dickey et al. also demonstrated that the skin oxide of eGaIn primarily consists of oxides of gallium which is consistent with the fact that gallium is highly reactive towards oxygen at room temperature [19]. In this study, $GaIn_{24.5}$ (weight percent, Ga 75.5%, In 24.5%) was used throughout all experiments and the oxidation mechanism as mentioned above has offered a good explanation for its surface oxidation. It is worth noting that suitable material is not limited to $GaIn_{24.5}$. After atomization, liquid alloy is discretized into small droplet and its surface is oxidized rapidly in the air. This allows it easier to adhere to the substrates and the surface oxide would guarantee the purity inside. With the decrease in size, the oxygen content of the droplets increases, finally causing an increase of oxygen in the deposited patterns. It was demonstrated that the adhesive force was stronger with the increase of oxygen. The work of adhesion of a liquid drop to a solid surface can be expressed by the Young-Dupre equation [23], i.e.

$$W=\gamma_L(1+\cos\theta) \quad (1)$$

where, $\gamma_L$ is the surface tension of the liquid and $\theta$ the contact angle. It can be concluded that a larger surface tension and a smaller contact angle will contribute to a stronger work of adhesion and the adhesion of the liquid to the solid will thus have better performance.

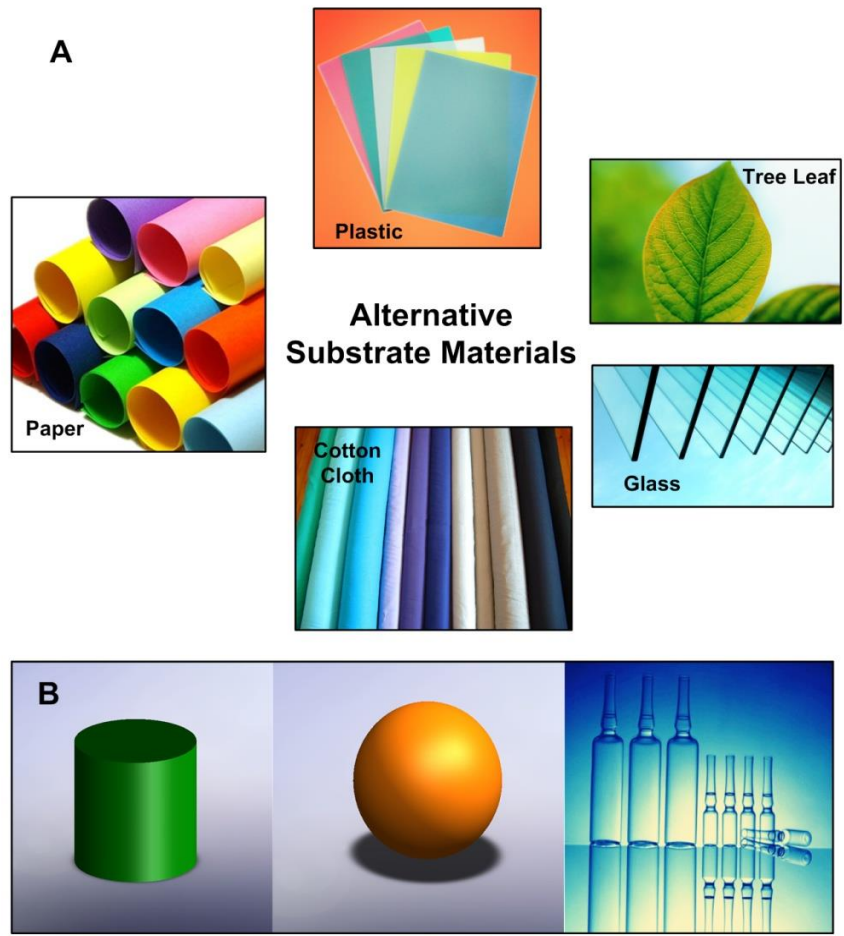

**Fig. 2** Generalized adaptability of liquid metal atomized spraying method for ubiquitous printed electronics. (A) Various typical material types; (B) Representative geometrical orientations of the target substrates.

To experimentally characterize the adhesion of the liquid metal ink, the surface tension of the droplets was measured by hanging drop method with a relative error of 0.15% and the contact angle was measured by five-point fitting based on the images with an accuracy of 0.5° (JC2000D3, POWEREACH). The $GaIn_{24.5}$ alloy was oxidized through constantly stirring in air at room temperature and the quantity of oxygen can be calculated by the weight increase of $GaIn_{24.5}$-based liquid metal with and without oxidization. As is observed (Fig. 1B), with the increase of oxygen content, the surface tension increases. However, an interesting phenomenon is that the change of the contact angle is not obvious: θ of a droplet without oxidation is 153.8 ° while θ of a droplet with the highest quantity of oxygen has a decrease of approximately 3 ° on the glass. All these data indicates an increase of W referring to Formula (1). Overall, the new method can synchronize the process of modification of liquid metal ink with the electronics fabrication on general substrates. It will bring about promising impacts in both energy field and circuit manufacturing technique in the coming time. To clearly outline the adaptability of the present method which also serves as experimental set up of the substrate types for spraying liquid metal, Fig. 2 illustrates various alternative substrate materials and multi-dimensional target orientations that can be printed with liquid metal electronics. The following work will demonstrate all these conceptual approaches which are expected to be significant in a wide variety of engineering situations.

## 3  Results

In this work, an airbrush is adopted for atomized spraying of liquid metal whose schematic structure is shown in Fig. 3A. Liquid metal is discharged through a fluid nozzle from the container and a hollow column of high-pressure air emitted from the air nozzle surrounding this liquid stream. The air produced by a pump will shred liquid metal fluid and make it disintegrated into droplets. A switch beside the container can control the amount of liquid metal. The whole process can be interpreted using Bernoulli's principle. High-purity liquid alloy $GaIn_{24.5}$ without deliberate oxidization is used during the whole experiment. Actually, a mask can be administrated to deposit specific patterns which bring about potential applications of a rapid prototyping of electronic component. However, it should be noted that the mask must be tightly positioned close to the substrates to gain printed electronics with clear boundaries. The location of the electronic devices can be reserved in advance where SMD (Surface Mounted Device) components will be mounted later. As it has been revealed that RTV silicone rubber can be adopted as isolating and packaging material to guarantee the functional stability of the circuit [12], therefore we mainly discuss the method of atomized spraying to deposit liquid metal electronics itself leaving the packaging issue for future discussion. A typical circuit made by the current rapid prototyping method based on atomized spraying of liquid metal droplets is presented in Fig. 3B and Fig. 3C and the application of the new method can be extended to more complex circuits and electronics such as crossed electrical wires and other crossed conductive parts.

At the initial stage, patterns were deposited onto the desired substrates ranging from rough to smooth surfaces. It has been demonstrated that the new method has a general applicability to fabricate liquid metal electronics on flexible substrate. The sprayed micro droplets show excellent wettability with almost any desired materials including plastic (Fig. 4A), polyvinylchloride (Fig. 4B), porous rubber (Fig. 4C), typing paper (Fig. 4D), cotton cloth (Fig. 4E), tree leaf (Fig. 4F) and more substrates which have very different surface roughness and material property. For the highly conductive liquid metal ink, as the sintering on the deposited material is not necessary throughout the whole process, the selection of substrates is thus no longer limited by the temperature. In this sense, the present method provides an alternative approach for making flexible, portable and foldable electronics. Fig. 4G gives the SEM graph surface morphology of conductive film printed by using GaIn$_{24.5}$ on glass and the diameter of the droplet ranges from approximately 700nm to 50μm. To gain a single droplet, the liquid metal was spayed one time rapidly on glass and the single droplet can be found on the edge of the spraying range.

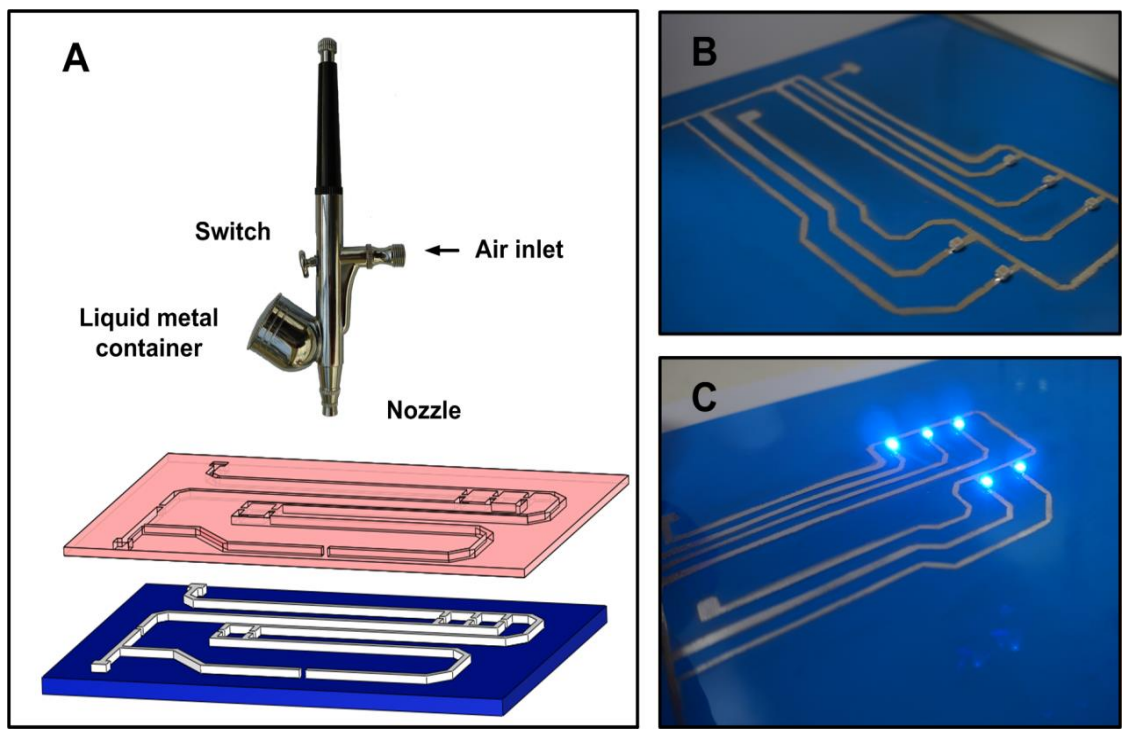

**Fig. 3** (A) Schematic of rapid prototyping circuits based on atomized spraying of liquid metal droplets; (B) and (C) present the printed typical liquid metal circuits on common plastic films (The LEDs' size is 3.5mm×2.77mm×1.93mm).

As it is small droplets deposited on the substrates, we speculate that such formed film may show interesting optical properties. This is because there is some space among liquid metal droplets which provides possibility for optical transmission path while they are connected with reserved electrical properties. All these features indicate that the method of liquid metal spraying offers a brand-new route for the direct fabrication of transparent conductive film. In this way, GaIn$_{24.5}$ can be deposited on the glass without additional complex treatment. For the present experiment, the sheet resistance of the deposited film was measured by four-probe method which always falls in milliohms scale per square with an optical transmittance below 40%. It is clear that the optical transmittance increases when the sheet resistance increases as there will be more porosity with fewer adjoining area of liquid metal. The sheet resistance Rs, defined as the resistance of identical width (w) and length (l) and uniform resistivity (ρ) and thickness (t), is expressed as Rs=ρ/t [24]. Fig. 5 depicts the optical transmittance of films with different sheet resistances which are measured by an UV visible spectrophotometer (UV-R928 PMT, Varian, Cary 5000) in the wavelength range from 300 to 800 nm. A transparent conductive film with an optical transmittance of 47% and a sheet resistance of 5.167Ω/□ is gained.

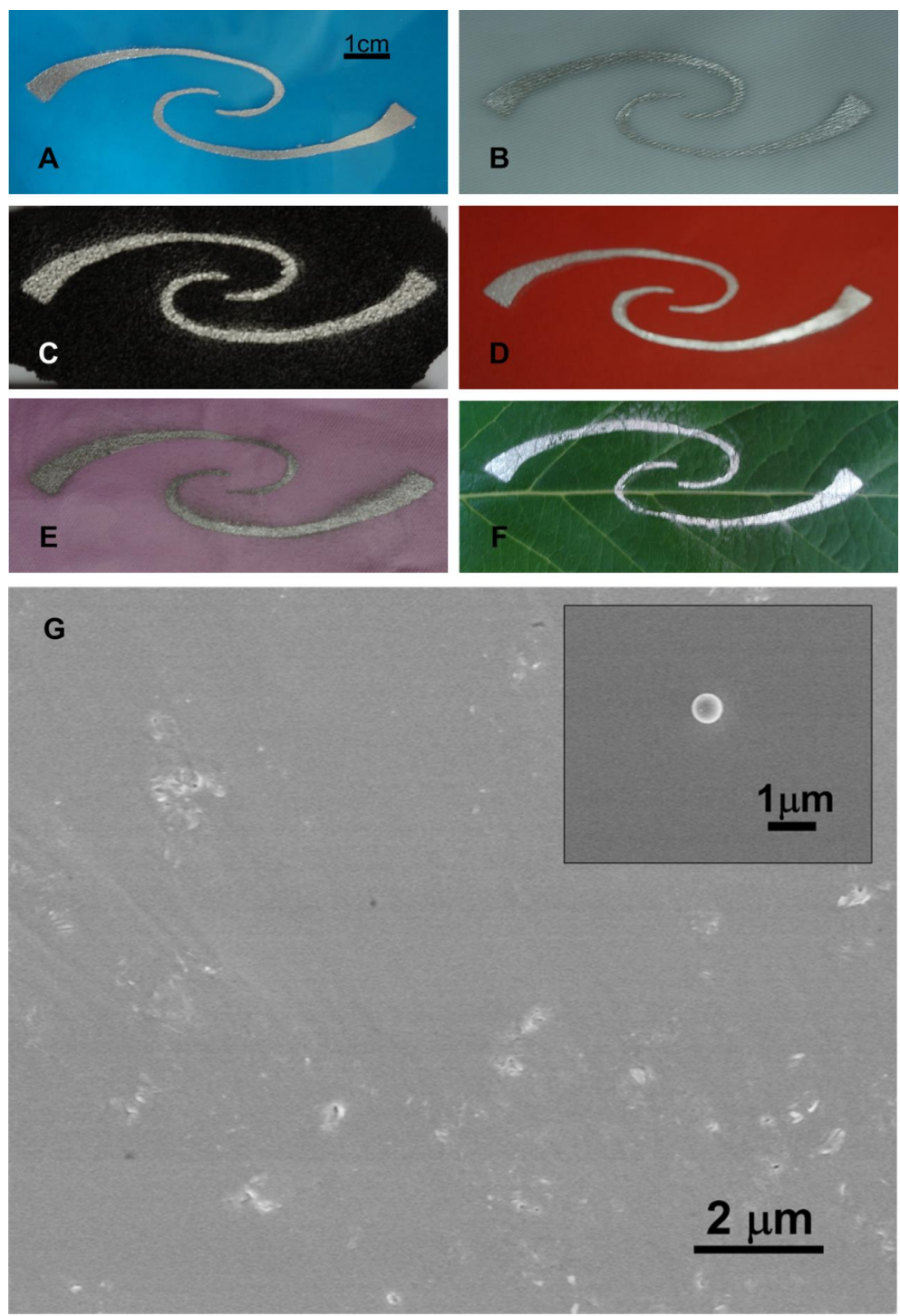

**Fig. 4** Demonstrated wettability of liquid metal droplets sprayed on different substrate materials. (A) Smooth polyvinylchlorid; (B) Rough polyvinylchlorid; (C) Porous rubber; (D) Typing paper; (E) Cotton cloth; (F) Tree leaf; (G) SEM graph of surface morphology of conductive film printed by using GaIn$_{24.5}$ through spraying and the inset shows one droplet sprayed out.

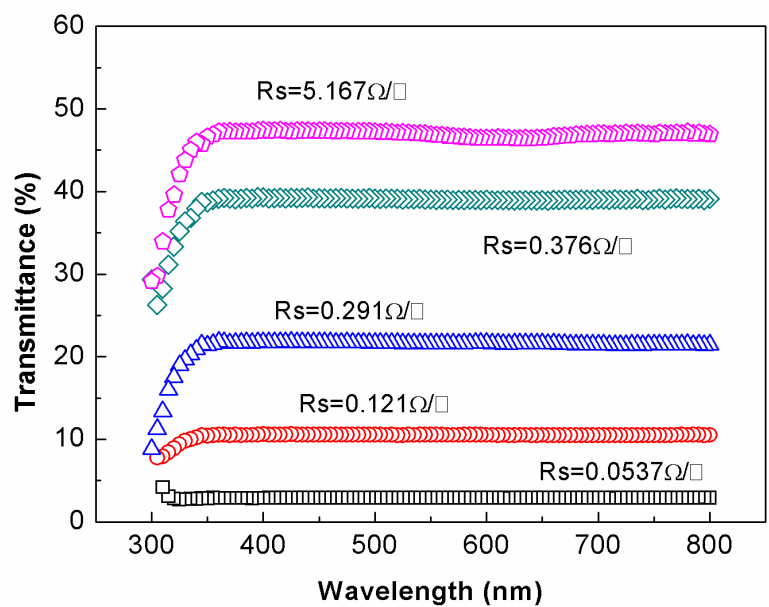

**Fig. 5** Optical transmittance of the porous liquid metal film with different sheet resistance.

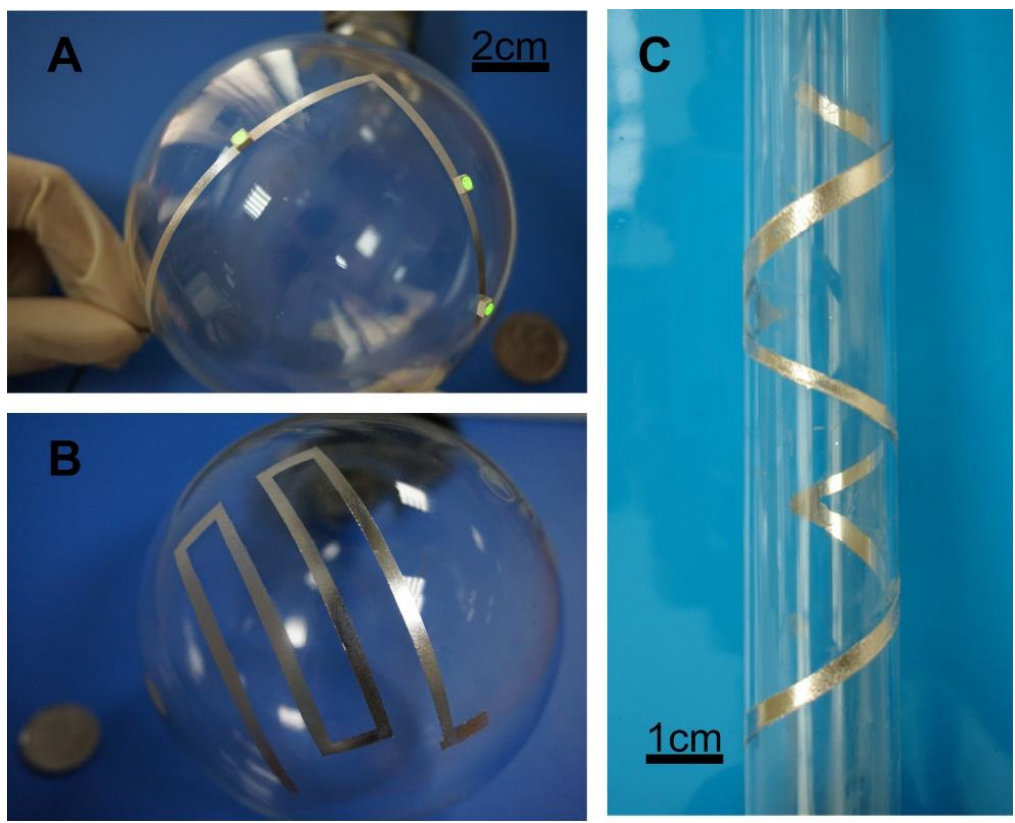

**Fig. 6** Demonstration experiments on the method of liquid metal deposition via spraying to different 3D surfaces. (A) Optical images of a spherical substrate display containing LED arrays or (B) electrical wires on the bottom of a flask. (C) Optical images of electrical wires on a cylindrical surface.

In addition, the new method can also be extended from 2D to 3D conformal electronics and Fig. 6 presents different conductive patterns deposited by spraying the liquid metal droplets on 3D substrates. This paves an innovative way for fabricating 3D functional electronics on various substrate materials to satisfy more actual demands in some special applications.

## 4  Discussion

Though the above experiments demonstrated the feasibility of the present method in a wide variety of printed electronics on different substrates ranging from rough to smooth surface and from 1D to 3D geometrical structures, there also opens spaces for further improvement. As in the current conceptual experiments, the airbrush is mainly held by hand to provide a simple and convenient way for routine use, there may have certain uncertainty regarding the position and height which would affect the objects' uniformity. To fulfill a more critical requirement in the future, the airbrush can be fixed on a bracket with controlled three-dimensional movement to realize the precise position of the liquid metal droplets.

Further, as addressed by the typical printing tasks as tackled in the present work, there exists a series of fundamental issues regarding the high performance printing and adhesion of the liquid metal droplets as well as material matching between the droplets and substrate. These need comprehensive characterization from micro scale to large size structure in the near future. It is important to determine and control the conditions of the droplets prior to the impingement for higher quality and homogeneity of the deposition in terms of microstructure and larger yield [25]. The droplet sizes from pneumatic nozzles are primarily a function of liquid flow rate, gas rate, atomizing gas pressure, nozzle diameters, flow ratio, and viscosity [26]. Fritsching [27] demonstrated that one was able to design a certain particle property in a wide range, such as the particle size distribution, particle shape and particle morphology, respectively. The previous work provides theoretical support for precise control of direct writing technology based on room temperature spraying. The mechanical properties of coatings by spraying depend on the shape of the splats formed from individual droplets as they impact together. Besides, splashing and incomplete flattening both degrade the coating quality since they may form voids in the deposit which then increases the porosity [28]. It is therefore essential to understand the impact dynamics of an individual liquid metal droplet. However, different from conventional droplets, liquid metal droplets have a portion of surface energy elastically stored in the skin of oxide [20].

More investigations should focus not only on the impact of liquid metal droplet on rigid plate, but also the impact on flexible substrate, spherical substrate and even metal liquid film. There may be opportunities to create electrical film with better uniformity through, for example, better design of nozzle structures. Other technologies, such as spinning technology and charging technology may help further improve the electronics fabrication quality for different applications. For example, molten metal droplets can be electrostatically charged and deflected onto a substrate with a measured accuracy of ±12.5 μm [11]. This also suggests the high resolution printed electronics via the present method. Before the new method can be put into large-scale application, some emerging problems such as how to improve the utilization way and how to recycle the remaining metal calls for further investigations in the coming time.

## 5 Conclusion

In summary, we have demonstrated that atomized spraying of liquid metal droplets can serve as a generalized way for ubiquitous printed electronics. As the small liquid metal droplets will be quickly oxidized when they are sprayed out, it is easy to adhere such metal inks to the desired target objects without stirring liquid metal for oxidation as a way to modify its compatibility with the substrates. The new method still maintains the unique merits as direct fabrication under room temperature and adaptability with different substrates. It further simplified the preparation process of liquid metal ink compared to the hand writing electronics. This opens a promising way for rapid electronics prototyping and will significantly reduce the time-to-market. RTV silicone rubber can be adopted as isolating and packaging material to guarantee the functional stability of the circuit. It is found from observation that liquid metal spraying method might make it possible to gain transparent conductive films. And a film with an optical transmittance of 47% and a sheet resistance of 5.167Ω/□ is obtained finally. Further investigation will focus on impact dynamics of liquid metal droplets and combination with other technologies, such as spinning and charging strategies, respectively. It is also expected that optimization on the method will yield the ability to create features with smaller size and higher quality. This finally would extend to the fabrication of novel devices such as soft resistors, inductors, capacitors, conductive wires, transparent conductive film with applications of stretchable electronics, skin circuit, flexible display and even photovoltaic devices.

**Acknowledgments**

The authors acknowledge valuable discussions with Prof. Yixin Zhou, Mr. Hainan Zhang, Ms. Lu Tian, Ms. Haiyan Li, and Mr. Meng Gao from Technical Institute of Physics and Chemistry (TIPC), Chinese Academy of Sciences.